\documentclass[english,twocolumn,aps,showpacs,floatfix]{revtex4}
\usepackage[T1]{fontenc}
\usepackage[latin1]{inputenc}
\usepackage{graphicx}
\usepackage{bm}
\usepackage{epsfig}
\usepackage{pslatex}
\usepackage[normalem]{ulem}  
\usepackage[dvips]{color}

\newcommand{\pt}{$p_T$ }

\newcommand{\pbar} {\mbox{$\overline{p}$}}

\def\Journal#1#2#3#4{{#1} {\bf #2}, #3 (#4)}

\def\PRL{ Phys. Rev. Lett.}

\makeatletter

\usepackage{babel}
\makeatother

\begin{document}

\title{Charm hadron production in relativistic heavy ion collisions within
a quark combination model}

\author{Tao Yao}

\affiliation{School of Physics, Shandong University, Jinan,
Shandong 250100, P. R. China}

\author{Wei Zhou}

\affiliation{School of Physics, Shandong University, Jinan,
Shandong 250100, P. R. China}

\author{Qu-Bing Xie}

\affiliation{School of Physics, Shandong University, Jinan, Shandong 250100,
P. R. China}




\date{\today}

\begin{abstract}
We investigate charm hadron production in relativistic heavy ion
collisions with the quark combination model. The $p_{T}$
dependencies of the charm baryon-to-meson ratios such as
$\frac{\Lambda_{c}+\bar{\Lambda}_{c}}{D^{0}+\bar{D}^{0}}$,
$\frac{\Lambda_{c}+\bar{\Lambda}_{c}}{D^{+}+D^{-}}$ and
$\frac{\Lambda_{c}+\bar{\Lambda}_{c}}{D_{s}^{+}+D_{s}^{-}}$ in Au+Au
collisions at 200 GeV are obtained. The charm baryon enhancement in
the intermediate $p_{T}$ range is very prominent, which further,
together with the strangeness enhancement, affects the charm hadron
ratios. The modified charm hadron ratios lead to a $\sim17\%$
increase of the charm cross section given by PHENIX. The
dependencies of the charm hadron ratios on energy, centrality, and
other parameters are also investigated. Predictions of the charm
hadron ratios for the upgrade of RHIC and for LHC are presented.
\end{abstract}




\maketitle

\section{Introduction}

Charm production in high energy heavy ion collisions is one of the
hot topics of both theory and experiment. Because of the large mass,
charm quarks are believed to be produced mainly via initial gluon
fusion in the early stage of relativistic heavy ion collisions
\cite{Lin:1994xma,Cacciari:2005rk}. They are also regarded as a
unique tool for probing the hot dense matter or quark-gluon
plasma(QGP) created in these collisions. For example, through the
charm quark energy loss \cite{c-energyloss}, charm flow
\cite{c-flow} and $J/\psi$ production (suppression or enhancement)
\cite{J/psipaper,Andronic:2003zv,Andronic:2006ky}, etc., one can
learn much about the QGP.

Since the startup of the BNL Relativistic Heavy Ion Collider (RHIC),
PHENIX and STAR collaborations have made many measurements on charm
production
\cite{Tai:2004bf,stardAucharm,Adler:2005ab,Abelev:2006db,phenixAuAu,
phenixpp,Zhong:2007iq,Baumgart:2007eu}. The binary scaling of the
total charm cross section $\sigma_{c\bar{c}}$ has been observed by
the two collaborations. However, they give quite different values of
the binary scaled charm cross section $\sigma_{c\bar{c}}^{NN}$.
There are two important ratios used for obtaining
$\sigma_{c\bar{c}}^{NN}$ in the experiments. One is
\begin{equation}\label{eq:defRec}
R_{e/c\bar{c}}\equiv N_{(e^{+}+e^{-})/2}/N_{c\bar{c}} =
\sigma_{(e^{+}+e^{-})/2}^{NN}/\sigma_{c\bar{c}}^{NN}
\end{equation}
used by PHENIX to convert the multiplicity or cross section of
nonphotonic electron into the charm cross section. Here
$N_{c\bar{c}}$ is the number of $c\bar{c}$ pairs created in the
collisions. Another ratio is
\begin{equation}\label{eq:defRdc}
R_{D^{0}/c\bar{c}}\equiv N_{(D^{0}+\bar{D}^{0})/2}/N_{c\bar{c}} =
\sigma_{(D^{0}+\bar{D}^{0})/2}^{NN}/\sigma_{c\bar{c}}^{NN}
\end{equation}
used by STAR in deriving $\sigma_{c\bar{c}}^{NN}$ from the yield of
$D^{0}$. $R_{e/c\bar{c}}$ \cite{Sorensen:2005sm} and
$R_{D^{0}/c\bar{c}}$ are both affected by the charm hadron ratios.

The enhancement of baryon production in the intermediate $p_{T}$
range observed in Au+Au reactions at RHIC suggests strongly a
coalescence/recombination (CO/RE) hadronization mechanism
\cite{co/repaper}. This mechanism can also result in an enhancement
of charm baryons, \footnotemark[1] \footnotetext[1]{Another
mechanism on $\Lambda_{c}$ enhancement is studied in Ref.
\cite{Lee:2007wr}.} so the charm hadron ratios and $R_{e/c\bar{c}}$,
$R_{D^{0}/c\bar{c}}$ in Au+Au collisions should be different from
those in $pp(\bar{p})$ reactions or $e^{+}e^{-}$ annihilations.
Unfortunately, these ratios cannot be obtained through theory
calculations model-independently, and it is now difficult to measure
them directly in experiments because of the difficulty of charm
hadron reconstruction in Au+Au collisions. \footnotemark[2]
\footnotetext[2]{It may be possible at RHIC after detector upgrades
\cite{upgrades}.} The two key ratios at RHIC are currently from the
$pp(\bar{p})$ reactions or $e^{+}e^{-}$ annihilations. Then one must
ask, how large are the corrections to $\sigma_{c\bar{c}}$ from the
two ratios in Au+Au collisions, and are the corrections able to
account for the discrepancy of $\sigma_{c\bar{c}}^{NN}$ measured by
the two collaborations?

A new method is proposed in Ref. \cite{Liu:2006id} to determine
$\sigma_{c\bar{c}}^{NN}$ by measuring the spectrum of nonphotonic
muon, so the ratio $R_{\mu/c\bar{c}}$, the inclusive branching ratio
to muons [($\mu^{+}+\mu^{-})/2$] in $AA$ reactions, is also
required.

The charm cross section is one of the most important issues of charm
physics in heavy ion collisions. According to Eqs. (\ref{eq:defRec})
and (\ref{eq:defRdc}), the accurate $\sigma_{c\bar{c}}^{NN}$
measurement depends on the accurate ratios $R_{e/c\bar{c}}$ and
$R_{D^{0}/c\bar{c}}$, so it is an important issue of charm physics
to determine $R_{e/c\bar{c}}$, $R_{D^{0}/c\bar{c}}$,
$R_{\mu/c\bar{c}}$ and the charm hadron ratios in $AA$ collisions.
However, in most combination models, the hadronic ``combination
function'' is necessary to obtain the yield of the hadron. The
combination function denotes the probability for (anti)quarks to
form a hadron and is determined by the hadron wave function. As the
wave functions of charm hadrons are unknown, it is difficult for
these models to study the issue quantitatively. In addition, these
models do not satisfy the unitarity \cite{unitarityinCORE} which is
important to the issue as well. By now, only one model
\cite{Andronic:2003zv,Andronic:2006ky} has predicted the
comprehensive charm hadron ratios at RHIC and LHC within statistical
framework. On the other hand, the quark combination model (QCM) we
developed satisfies the unitarity and has reproduced the global
properties \cite{Xie:1988wi,Xie:1997ap,epSDQCM,$AA$SDQCM} of SU(3)
hadrons without the explicit combination functions. Based on the
success of QCM, we further extend it to SU(4) flavor symmetry, and
QCM is then suitable for studying quantitatively the charm hadron
production in heavy ion collisions.

In this paper, we quantitatively study the effects on the charm
hadron ratios from the baryon enhancement and the strangeness
enhancement in $AA$ reactions within the QCM. We find the charm
hadron ratios and the key ratios $R_{e/c\bar{c}}$,
$R_{\mu/c\bar{c}}$ are substantially different from those in the
$pp(\bar{p})$ reactions or $e^{+}e^{-}$ annihilations. Their
dependencies on energy, centrality, and some parameters are all
investigated. The parameters include the yield ratio of the primary
charm vector meson to the pseudoscalar meson $V_{c}/P_{c}$, the
yield ratio of primary charm decuplet baryon to octet baryon
$D_{c}/O_{c}$, the quark number ratio $N_{\bar{d}}/N_{d}$, the
strangeness suppression factor $\lambda_{s}$, and so on. Here
$\lambda_{s}$ denotes the number ratio of newly produced strange
quarks to $u(d)$ quarks, $\lambda_{s} \equiv
\frac{2N_{s\bar{s}}}{N_{u\bar{u}}+N_{d\bar{d}}}=
\frac{N_{\bar{s}}}{N_{\bar{u}}}=\frac{N_{\bar{s}}}{N_{\bar{d}}}$.
For the first time in combination or CO/RE mechanism,
\footnotemark[3] \footnotetext[3]{Note that the combination picture
or CO/RE mechanism is the same in all
coalescence/recombination/combination models, but the methods to
implement the mechanism are different in detail.} the extensive
charm hadron ratios and the total branching ratios $R_{e/c\bar{c}}$,
$R_{D^{0}/c\bar{c}}$, $R_{\mu/c\bar{c}}$ in $AA$ collisions are
predicted at RHIC and LHC. Since the dealings with hadronization are
different, it is not surprising that some predictions are different
from those in the statistical hadronization model
\cite{Andronic:2003zv,Andronic:2006ky}. The predictions can be
examined and the different hadronization mechanisms for charm
hadrons can be tested in future experiments.

\section{Quark combination model}

Our quark combination model (QCM) proposed some time ago
\cite{Xie:1987ds,Xie:1988wi,Xie:1997ap,epSDQCM,excSDQCM} demands
that quark(s) and/or antiquark(s) which are nearest in rapidity
combine into a hadron. It has been shown
\cite{Xie:1988wi,Xie:1997ap} that such a demand is in agreement with
the fundamental requirement of QCD and uniquely determines the quark
combination rule in the hadronization process. QCM has been
successfully applied to $e^{+}e^{-}$ annihilations and $pp(\bar{p})$
collisions \cite{Xie:1988wi,Xie:1997ap,epSDQCM}. Recently, we have
extended it to the RHIC reactions and have reproduced the global
properties of hadrons such as hadronic multiplicities, $p_{T}$
spectra, elliptic flows, and rapidity distributions
\cite{$AA$SDQCM}.

QCM is particularly designed for describing the hadronization in the
quark CO/RE scheme where the properties of the constituent
(anti)quarks before hadronization are taken as inputs. In this work,
the charm quarks are supposed to be distributed randomly in the
light (anti)quark sea after the QGP evolution, and all the
(anti)quarks, including charm quarks, combine into hadrons within
QCM. All the ground state hadrons, namely, 120-plet baryons and
64-plet mesons for SU(4) quarks, are considered systematically. The
higher excited states are not included. The final state hadrons are
obtained after all resonances are treated through the decay
subroutine in PYTHIA6.4.16 \cite{Sjostrand:2000wi}. Some details of
QCM can also be seen in Ref. \cite{$AA$SDQCM}.

We determine the parameters $\lambda_{s}$ and $N_{\bar{d}}/N_{d}$ by
fitting the two hadron ratios $K^{+}/\pi^{+}$ (or $K^{-}/\pi^{-}$)
and $\pbar/p$. The other hadron ratios can be obtained accordingly
with QCM. The parameter $N_{\bar{d}}/N_{d}$, a measure of the net
baryon number in QGP, is taken as 0.859, 0.915, and 0.927 for
$\sqrt{s_{NN}}= 62.4, 130$, and 200 GeV respectively. Likewise,
$\lambda_{s}$ is 0.52, 0.48, and 0.48 for the three RHIC energies,
and this is consistent with the data fitting in Ref.
\cite{Becattini:2008yn}. Similar to the same reference, we suppose
$\lambda_{s}$ is saturated in $AA$ reactions above 200 GeV, that is,
QGP created at LHC also has $\lambda_{s}= 0.48$. Assuming the
parameters $V_{c}/P_{c}$ and $D_{c}/O_{c}$ are universal in
relativistic $AA$ reactions, we use 3.0 and 0.5 as their default
values in QCM at various energies, respectively. The same default
value of $V_{c}/P_{c}$ is also used in PYTHIA.

For simplicity and to reduce the uncertainty of the input, same as
Ref. \cite{Sorensen:2005sm}, the bottom quarks are not considered in
our calculations. According to the pQCD prediction in Ref.
\cite{Cacciari:2005rk}, the bottom effects on the spectrum of
nonphotonic electron are mainly manifested in a higher $p_{T}$ range
($p_{T}>4$ GeV), so we can reliably study the charm hadron
multiplicities or ratios while neglecting the bottom contributions.
We will not discuss charmonia production here, but will study it in
a future paper, as this issue is more complicated and still under
debate.

\section{Results and discussions}\label{sec:Results}

\subsection{$p_{T}$ dependencies of charm baryon-to-meson ratios}

\begin{figure}
\includegraphics[width=0.45\textwidth, height=0.4\textwidth]{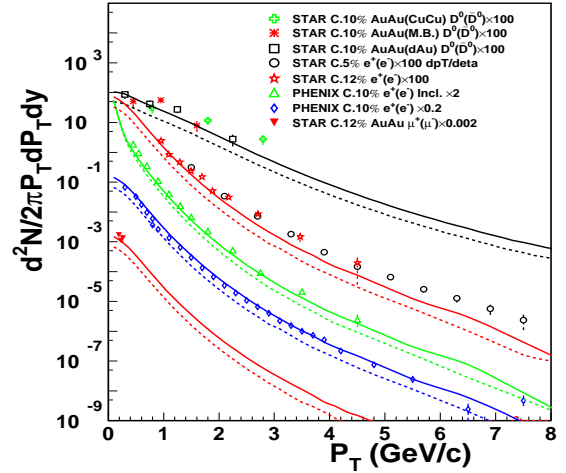}
\caption{(Color online) Midrapidity $p_{T}$ spectra at 200 GeV for
$D^{0}(\bar{D^{0}})$ meson, inclusive electron, nonphotonic
electron, and muon. The solid (dashed) lines are from
$\sigma_{c\bar{c}}^{NN}$ by STAR (PHENIX). $D^{0}$ data tagged with
(CuCu) or ($d$Au) are obtained from Cu+Cu or $d$+Au collisions based
on the binary scaling. Data are from Refs.
\cite{Baumgart:2007eu,Zhong:2007iq,stardAucharm,Abelev:2006db,phenixAuAu}.}
\label{cap:oncharms}
\end{figure}

Since most of the data available on charm production in Au+Au
collisions are at $\sqrt{s_{NN}}=200$ GeV, we first study charm
production in the top central Au+Au collisions at this energy.
According to the experimental observation of binary scaling of the
nonphotonic electron spectra in low $p_{T}$ range, we assume that
the total charm cross section or $N_{c\bar{c}}$ in relativistic $AA$
collisions is proportional to the nucleon-nucleon collision number
$N_{bin}$. The assumption is consistent with the point that charm
quarks are mostly produced via primary hard scattering, and the
number of $c\bar{c}$ pairs $N_{c\bar{c}}$, as input in QCM, is given
by
\begin{equation}
N_{c\bar{c}}= \langle N_{bin} \rangle \sigma_{c\bar{c}}^{NN}/\sigma_{inel}^{pp}.
\label{eq:Ncc}
\end{equation}
Here the Au+Au charm cross section per nucleon-nucleon collision
$\sigma_{c\bar{c}}^{NN}=1.4$ mb \cite{stardAucharm}, the $pp$
inelastic cross section $\sigma_{inel}^{pp}=42$ mb and $\langle
N_{bin} \rangle=1051 (939.5)$ \cite{Adams:2004cb} are adopted. Then
the total $N_{c\bar{c}}$ is about 35 (31) at 0-5\% (0-10\%)
centrality.

The charm quark spectrum that was input as a normalized form
$f_{c}(p_{T}) = (1.0+3.185p_{T}^{2})^{-2.7}/0.354$ is extracted by
fitting the STAR spectra of $D^{0}(\bar{D}^{0})$ and electron. The
fitted spectra in Fig. \ref{cap:oncharms} are slightly softer than
the STAR data considering no bottom contributions. The solid lines
are from the STAR $\sigma_{c\bar{c}}^{NN}$, while the dashed lines,
all lower than the data, correspond to the PHENIX
$\sigma_{c\bar{c}}^{NN} = 0.622$ mb \cite{Adler:2004ta}. Hereafter,
the STAR $\sigma_{c\bar{c}}^{NN}$ 1.4 mb is used in all calculations
at 200 GeV.

\begin{figure}
\includegraphics[scale=0.4]{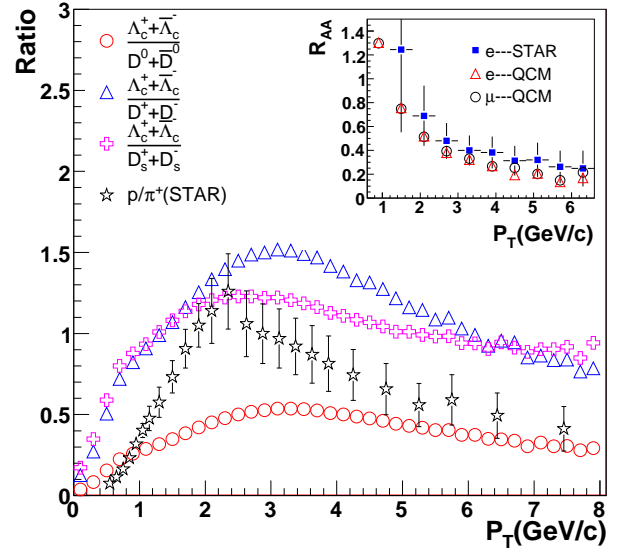}
\caption{(Color online) Midrapidity $p_{T}$ dependencies for the
ratios of charm baryons to mesons at 200 GeV. Inset shows $R_{AA}$
of nonphotonic lepton from QCM.} \label{cap:charmB-M_pt}
\end{figure}

The $p_{T}$ dependencies of the charm baryon-to-meson ratios
$\frac{\Lambda_{c}+\bar{\Lambda}_{c}}{D^{0}+\bar{D}^{0}}$,
$\frac{\Lambda_{c}+\bar{\Lambda}_{c}}{D^{+}+D^{-}}$, and
$\frac{\Lambda_{c}+\bar{\Lambda}_{c}}{D_{s}^{+}+D_{s}^{-}}$ are
plotted in Fig. \ref{cap:charmB-M_pt}. It is clear to see that the
charm baryon enhancement in intermediate $p_{T}$ range is very
prominent, which is similar to that of the $p/\pi$ ratio. Note that
the peak of $\frac{\Lambda_{c}+\bar{\Lambda}_{c}}{D^{+}+D^{-}}$ is
even higher than
$\frac{\Lambda_{c}+\bar{\Lambda}_{c}}{D_{s}^{+}+D_{s}^{-}}$. This is
because the yield of $D_{s}^{+}(D_{s}^{-})$ is larger than
$D^{+}(D^{-})$ around $p_{T}=3$ GeV/$c$ due to the decay effect. The
shapes of the $p_{T}$ dependencies of these ratios are dependent on
the CO/RE mechanism, decay effect, and the charm quark spectrum
obtained from fitting data shown in Fig. \ref{cap:oncharms}.
Comparing with the data of $p/\pi^{+}$ \cite{Abelev:2006jr} in the
same Au+Au collisions, one can see that all peaks broaden and shift
to the right because essentially the spectrum of the charm quark is
much harder than that of the $u(d)$ quark.

The enhancement can certainly lead to the $R_{AA}$ enhancement of
charm baryons ($R_{AA} \equiv \frac{dN_{AA}/dyd^{2}p_{T}} {\langle
N_{bin}\rangle dN_{pp}/dyd^{2}p_{T}}$) in the same $p_{T}$ region.
It can also result in a suppression of nonphotonic lepton spectrum
and thus a substantial suppression to its nuclear modification
factor $R_{AA}$ in the intermediate $p_{T}$ range
\cite{Sorensen:2005sm, MartinezGarcia:2007hf}. This is just the part
of $R_{AA}$ of nonphotonic lepton that arose from the CO/RE
hadronization process other than the quark energy loss. In the inset
of Fig. \ref{cap:charmB-M_pt}, our calculations also predict a very
similar $R_{AA}$ of the nonphotonic muon to that of the nonphotonic
electron. \footnotemark[4] \footnotetext[4]{The nonphotonic lepton
spectra in $pp$ reactions are obtained from PYTHIA6.2 with
parameters given in \cite{Lin:2006rp}.}

Within the CO/RE hadronization mechanism, we discuss the $R_{AA}$
ordering with strangeness content for charm hadrons in heavy ion
collisions. In fact, the $R_{AA}$($R_{cp}$) ordering with
strangeness content for strange hadrons has been observed at RHIC
\cite{Salur:2005nj, Abelev:2007rw}, which is a natural result in the
CO/RE mechanism for the strangeness enhancement in heavy ion
collisions. The effect should also be manifested by the strange
charm hadrons such as $R_{AA}$ of $D_{s}^{+}(D_{s}^{-})$ in
comparison with that of nonstrange charm mesons, $R_{AA}$ of
$\Xi_{c}$ with respect to that of $\Lambda_{c}^{+}$ and so on. The
$R_{AA}$ ordering with strangeness content for charm hadrons is also
a powerful signal for proving the CO/RE hadronization mechanism for
open charm hadrons. The precise measurement of it in future
experiments can help clarify various hadronization mechanisms.

\subsection{Ratios in $AA$ collisions at 200 GeV}

As discussed in the last subsection, the CO/RE mechanism results in
the charm baryon enhancement which can further affect the charm
hadron ratios in relativistic $AA$ reactions. In this subsection,
the effects on charm hadron ratios and the three key ratios are
quantitatively studied within the QCM.

The rapidity densities of single-charm hadrons and their ratios in
Au+Au collisions at 200 GeV are listed in Table
\ref{tab:densityratios}. We can see that the rapidity densities for
charm hadrons are different from those for charm antihadrons because
of the net $u/d$ quarks in QGP. The yield of $D^{0}(\bar{D}^{0})$
obtained at midrapidity $dN_{(D^{0}+\bar{D}^{0})/2)}/dy|_{y=0}
\approx 3.1$ is in agreement with the STAR data from the $d$+Au
reactions \cite{stardAucharm} assuming a binary scaling of
$N_{c\bar{c}}$. In Ref. \cite{Adler:2004ta}, PHENIX modified the
PYTHIA default charm hadron ratios, and they obtain
$R_{e/c\bar{c}}=9.5\pm0.4\%$ by using $D^{+}/D^{0}=0.45\pm0.1$,
$D_{s}^{+}/D^{0}=0.25\pm0.1$, and
$\Lambda_{c}^{+}/D^{0}=0.1\pm0.05$. Obviously, $D_{s}^{+}/D^{+}$
from the PHENIX ratios is $0.25/0.45\approx0.56$. These ratios are
apparently different from those in $AA$ collisions listed in Table
\ref{tab:densityratios}. Especially, the ratio
$\Lambda_{c}^{+}/D^{0}$ is much enhanced in CO/RE mechanism in $AA$
reactions, as can also be seen in Fig. \ref{cap:charmB-M_pt}. It is
also impressive that the ratio $D_{s}^{+}/D^{+}$ is almost equal to
1, which is much higher than that (0.56) used by PHENIX. The first
reason is that $\lambda_{s}$ is taken as 0.48, much larger than that
in $pp$ reactions. The second reason is that $D_{s}^{\ast\pm}$
totally decays to $D_{s}^{\pm}$, but only about 32.3$\%$ of the
$D^{\ast\pm}$ particles decay to $D^{\pm}$ mesons, and the other
67.7$\%$ decay to $D^{0}(\bar{D}^{0})$ mesons. The last reason is
that the default value of $V_{c}/P_{c}$ is 3. However, even if
$V_{c}/P_{c}$ equals 1.5, the ratio $D_{s}^{+}/D^{+} \approx
\frac{0.48\times(1.5+1)}{1.5\times32.3\%+1} \approx 0.81$ is still
higher than 0.56.

Considering the effect from net baryon number, the value of
$R_{e/c\bar{c}}$ at midrapidity should be calculated by

\begin{widetext}
\begin{eqnarray}
R_{e/c\bar{c}}&=&\frac{6.87^{+0.28}_{-0.28}\%\times(1+\frac{\bar{D}^{0}}{D^{0}})+17.2^{+1.9}_{-1.9}\%\times
(\frac{D^{+}}{D^{0}}+\frac{D^{-}}{D^{0}})+8^{+6}_{-5}\%\times(\frac{D_{s}^{+}}{D^{0}}+
\frac{D_{s}^{-}}{D^{0}})+4.5^{+1.7}_{-1.7}\%\times(\frac{\Lambda_{c}^{+}}{D^{0}}+
\frac{\Lambda_{c}^{-}}{D^{0}})}{1+\frac{\bar{D}^{0}}{D^{0}}+\frac{D^{+}}{D^{0}}+
\frac{D^{-}}{D^{0}}+\frac{D_{s}^{+}}{D^{0}}+\frac{D_{s}^{-}}{D^{0}}+
\frac{\Lambda_{c}^{+}}{D^{0}}+\frac{\Lambda_{c}^{-}}{D^{0}}}\nonumber\\
&=&\frac{6.87^{+0.28}_{-0.28}\%\times(D^{0}+\bar{D}^{0})+17.2^{+1.9}_{-1.9}\%\times
(D^{+}+D^{-})+8^{+6}_{-5}\%\times(D_{s}^{+}+D_{s}^{-})+4.5^{+1.7}_{-1.7}\%\times(\Lambda_{c}^{+}+
\Lambda_{c}^{-})}{(D^{0}+\bar{D}^{0})+(D^{+}+D^{-})+(D_{s}^{+}+D_{s}^{-})+(\Lambda_{c}^{+}+
\Lambda_{c}^{-})}.
\label{eq:Rec}
\end{eqnarray}
\end{widetext}
These charm hadron ratios in Au+Au collisions give a lower
$R_{e/c\bar{c}} = 8.48^{+1.05}_{-0.90}\%$ than that used by PHENIX,
and it can lead to $\sim12\%$ increase of $\sigma_{c\bar{c}}^{NN}$
based on the PHENIX nonphotonic electron data. Here the errors of
$R_{e/c\bar{c}}$ are from the branching ratio uncertainties. On the
PDGlive web-page, the newest branching ratio for $ D^{0} (D^{+})
\rightarrow e^{+} + anything $ is $6.53\pm0.17\%$ ($16.0\pm0.4\%$),
from which the updated $R_{e/c\bar{c}} = 8.10^{+0.99}_{-0.84}\%$
results in $\sim17\%$ enhancement of $\sigma_{c\bar{c}}^{NN}$
(central value). That is, the ratio $R_{e/c\bar{c}}$ used by PHENIX
decreases $\sigma_{c\bar{c}}^{NN}$ by $\sim17\%$ and hence enlarges
the difference of $\sigma_{c\bar{c}}^{NN}$ between STAR and PHENIX.

Strictly speaking, the ratio obtained by Eq. (\ref{eq:Rec}) may be
different from the `real' $R_{e/c\bar{c}}$ ratio, which can be
calculated directly in QCM from the $R_{e/c\bar{c}}$ definition of
Eq. (\ref{eq:defRec}), i.e., from the yield of nonphotonic electrons
$N_{(e^{+}+e^{-})/2}|_{y=0}$ and the number of charm quark pairs
$N_{c\bar{c}}|_{y=0}$ at midrapidity. Note that the denominator of
Eq. (\ref{eq:Rec}) is smaller than $N_{c\bar{c}}|_{y=0}$, and the
numerator of Eq. (\ref{eq:Rec}) can also be different from
$N_{(e^{+}+e^{-})/2}|_{y=0}$. The value from Eq. (\ref{eq:defRec})
becomes $8.33\%$, smaller than $8.48\%$, and this leads to about a
2$\%$ enhancement (systematic error) of $\sigma_{c\bar{c}}^{NN}$.
Based on our calculations, such systematic errors for
$R_{e/c\bar{c}}$ at the various energies studied in the paper are
all smaller than $0.5\%$, and the resulting enhancements of
$\sigma_{c\bar{c}}^{NN}$ are all not more than $5\%$.

\begin{table}
\caption{Midrapidity densities of open charm hadrons and their
ratios for top central collisions at $\sqrt{s_{NN}}=62.4, 130, 200$
and 5500 GeV. $R_{e/c\bar{c}}$ and $R_{D^{0}/c\bar{c}}$ are from
Eqs. (\ref{eq:Rec}) and (\ref{eq:Rdc}).} \label{tab:densityratios}
\begin{tabular}{c|c|c|c|c|c|c|c|c} \hline
$dN/dy$ & 62.4 & 130  &  200 & Ratios & 62.4 & 130 & 200 & 5500 \\
\hline
$D^{+}$  & 0.217 & 0.532 & 0.938 & $D^{+}/D^{0}$ & 0.317 & 0.317 & 0.317  & 0.318 \\
$D^{-}$   & 0.252 & 0.582 & 1.011 & $D_{s}^{+}/D^{0}$ & 0.336 & 0.309 & 0.310 & 0.311 \\
$D^{0}$   & 0.684 & 1.678 & 2.954 & $\Lambda_{c}^{+}/D^{0}$ & 0.353 & 0.311 & 0.301 & 0.253 \\
$\bar{D}^{0}$  & 0.787 & 1.820 & 3.168 & $\bar{D}^{0}/D^{0}$ & 1.151 & 1.085 & 1.073 & 1.005   \\
$D_{s}^{+}$  & 0.229 & 0.519 & 0.915 & $D^{-}/D^{0}$ & 0.369 & 0.347 & 0.342 & 0.319 \\
$D_{s}^{-}$  & 0.229 & 0.520 & 0.915 & $D_{s}^{-}/D^{0}$ & 0.335 & 0.310 & 0.310 & 0.311 \\
$\Lambda_{c}^{+}$    & 0.241 & 0.522 & 0.890 & $\bar{\Lambda}_{c}^{-}/D^{0}$ & 0.205 & 0.227 & 0.230 & 0.249 \\
$\bar{\Lambda}_{c}^{-}$ & 0.140& 0.381 & 0.679 & $\Sigma_{c}^{+}/D^{0}$ & 0.081 & 0.071 & 0.069 & 0.057  \\
$\Sigma_{c}^{+}$  & 0.055 & 0.120 & 0.203 & $\Sigma_{c}^{0}/D^{0}$ & 0.076 & 0.066 & 0.064 & 0.052   \\
$\bar{\Sigma}_{c}^{-}$  & 0.032& 0.087 & 0.155 & $\Xi_{c}^{+}/D^{0}$ & 0.078 & 0.068 & 0.067 & 0.062 \\
$\Sigma_{c}^{0}$  & 0.052 & 0.111 & 0.188 & $\Xi_{c}^{0}/D^{0}$ & 0.079 & 0.068 & 0.067 & 0.062 \\
$\bar {\Sigma}_{c}^{0}$  & 0.029 & 0.080 & 0.142 & $D_{s}^{+}/D^{+}$ & 1.058 & 0.975 & 0.976 & 0.980 \\
$\Xi_{c}^{0}$  & 0.054 & 0.115 & 0.198 & $\bar{\Lambda}_{c}^{-}/\Lambda_{c}^{+}$ & 0.581 & 0.730 & 0.763 & 0.982 \\
$\Xi_{c}^{+}$  & 0.053 & 0.113 & 0.197 & $\Sigma_{c}^{+}/\Lambda_{c}^{+}$ & 0.228 & 0.230 & 0.228 & 0.226 \\
$e_{c}^{\pm}$  & 0.123 & 0.295 & 0.520 & $R_{e/c\bar{c}}$ & 0.085 & 0.085 & 0.085 & 0.085  \\
$\mu_{c}^{\pm}$  & 0.125 & 0.298 & 0.526 & $R_{D^{0}/c\bar{c}}$ & 0.529 & 0.534 & 0.534 & 0.532 \\
$N_{c\bar{c}}$ & 1.512 & 3.560 & 6.240 & $R_{\mu/c\bar{c}}$ & 0.083 & 0.084 & 0.084 & 0.086 \\
\hline
\end{tabular}
\end{table}

Another ratio used by STAR in Refs. \cite{stardAucharm,
Baumgart:2007eu} is $R_{D^{0}/c\bar{c}}=0.54\pm0.05$ from
$e^{+}e^{-}$ annihilation data at 91 GeV. We can obtain the ratio
approximately through the charm hadron ratios at midrapidity, that
is,
\begin{eqnarray}
R_{D^{0}/c\bar{c}}&=&\frac{1+\frac{\bar{D}^{0}}{D^{0}}}{1+\frac{\bar{D}^{0}}{D^{0}}+\frac{D^{+}}{D^{0}}+
\frac{D^{-}}{D^{0}}+\frac{D_{s}^{+}}{D^{0}}+\frac{D_{s}^{-}}{D^{0}}+
\frac{\Lambda_{c}^{+}}{D^{0}}+\frac{\Lambda_{c}^{-}}{D^{0}}}\nonumber\\
&=&\frac{D^{0}+\bar{D}^{0}}{D^{0}+\bar{D}^{0}+D^{+}+D^{-}+D_{s}^{+}+D_{s}^{-}+\Lambda_{c}^{+}+
\Lambda_{c}^{-}}. \label{eq:Rdc}
\end{eqnarray}
Then we have 0.534 in QCM at most central collisions at 200 GeV. The
value is by chance very close to that used by STAR and leads to only
$\sim1\%$ correction of $\sigma_{c\bar{c}}^{NN}$.

If we calculate $R_{D^{0}/c\bar{c}}$ via its definition of Eq.
(\ref{eq:defRdc}), it changes to 0.491, which means $\sim8\%$
enhancement (systematic error) of $\sigma_{c\bar{c}}^{NN}$. As the
denominator $N_{c\bar{c}}|_{y=0}$ of Eq. (\ref{eq:defRdc}) is larger
than that of Eq. (\ref{eq:Rdc}), this kind of systematic error of
$R_{D^{0}/c\bar{c}}$ always enlarges $\sigma_{c\bar{c}}^{NN}$. The
errors of $R_{D^{0}/c\bar{c}}$ at different energies studied in the
paper are not more than $5\%$, and the corresponding maximal
increase of $\sigma_{c\bar{c}}^{NN}$ is $\sim10\%$.

Similarly, we calculate $R_{\mu/c\bar{c}}$ by
\begin{equation}\label{eq:Rmuc-y}
R_{\mu/c\bar{c}}=\frac{N_{(\mu^{+}+\mu^{-})/2}^{c}|_{y=0}}{N_{c\bar{c}}|_{y=0}},
\end{equation}
and 8.42\% is obtained in midrapidity range.

In short, the effect on $R_{e/c\bar{c}}$ in relativistic $AA$
collisions leads to $\sim17\%$ increase of $\sigma_{c\bar{c}}^{NN}$
measured by PHENIX, while the correction of $\sigma_{c\bar{c}}^{NN}$
by STAR from $R_{D^{0}/c\bar{c}}$ is only $\sim1\%$. The
modifications of $R_{e/c\bar{c}}$ and $R_{D^{0}/c\bar{c}}$ in heavy
ion collisions reduce the discrepancy of $\sigma_{c\bar{c}}^{NN}$
between STAR and PHENIX but are not enough to account for it.

\subsection{Parameter dependencies of the ratios}

Considering the uncertainties of some parameters in $AA$ reactions,
especially at LHC energies, it is important to know to which
parameters these charm hadron ratios and the three key ratios are
sensitive. The parameters $V_{c}/P_{c}$, $D_{c}/O_{c}$,
$\lambda_{s}$, $N_{\bar{d}}/N_{d}$, and the charm quark $p_{T}$
spectrum are all possible candidates for affecting these ratios. We
will study the effects from these factors respectively.

Based on the data in Refs. \cite{Gladilin:1999pj, Yao:2006px}, there
is still much uncertainty regarding $V_{c}/P_{c}$. We vary the value
of $V_{c}/P_{c}$ with other conditions unchanged to investigate its
effect on these ratios. The results are given in Table
\ref{tab:ratios-VP}. One can see that the ratio $\bar{D}^{0}/D^{0}$
is almost independent of $V_{c}/P_{c}$, and $R_{D^{0}/c\bar{c}}$
increases with $V_{c}/P_{c}$, whereas the other ratios all decrease
monotonically with it. This is because $\sim67.7\%$ of $D^{\ast\pm}$
decays to $D^{0}$($\bar{D}^{0}$), so more $D^{\ast\pm}$ will result
in a larger $D^{0}$ part of the total charm hadrons. Note that for
the $V_{c}/P_{c}$ variance from 0.5 to 3.0, both $R_{e/c\bar{c}}$
and $R_{\mu/c\bar{c}}$ decrease by only $\sim1\%$, that is, they are
not sensitive to $V_{c}/P_{c}$. The reason is that the lepton
branching ratios from charm hadrons are very small; however, the
$\sim1\%$ variance will lead to $\sim10\%$ correction of
$\sigma_{c\bar{c}}^{NN}$. In contrast, the increase of
$R_{D^{0}/c\bar{c}}$ is $\sim10\%$, corresponding to $\sim20\%$
variance of $\sigma_{c\bar{c}}^{NN}$.

We also study the effect of the parameter $D_{c}/O_{c}$ on these
ratios. The ratios, such as $ \Lambda_{c}^{\ast+}/\Lambda_{c}^{+}$,
vary with $D_{c}/O_{c}$ certainly, but all the ratios listed in
Table \ref{tab:ratios-VP} are hardly affected by it, as the charm
decuplet baryons almost totally transform into charm octet baryons.

\begin{table}
\caption{Dependencies of charm hadron ratios on the parameter
$V_{c}/P_{c}$, calculated at midrapidity for central (0-5\%) Au+Au
collisions at 200 GeV. $R_{e/c\bar{c}}$ and $R_{D^{0}/c\bar{c}}$ are
from Eqs. (\ref{eq:Rec}) and (\ref{eq:Rdc}). The symbol $\nearrow$
($\searrow$) denotes increase (decrease) with $V_{c}/P_{c}$.}
\begin{tabular}{c|c|c|c|c|c|c|c} \hline
$V_{c}/P_{c}$ & 0.5 & 1.0 & 1.5 & 2.0 & 2.5 & 3.0 & \\
\hline
$D^{+}/D^{0}$ & 0.620 & 0.482 & 0.411 & 0.367 & 0.339 & 0.317 & $\searrow$ \\
$D_{s}^{+}/D^{0}$ & 0.381 & 0.348 & 0.332 & 0.322 & 0.315 & 0.310 & $\searrow$\\
$\Lambda_{c}^{+}/D^{0}$ & 0.379 & 0.343 & 0.326 & 0.314 & 0.307 & 0.301 & $\searrow$\\
$\bar{D}^{0}/D^{0}$ & 1.072 & 1.071 & 1.071 & 1.073 & 1.074 & 1.073 & \\
$D^{-}/D^{0}$ & 0.670 & 0.520 & 0.443 & 0.396 & 0.366 & 0.342 & $\searrow$\\
$D_{s}^{-}/D^{0}$ & 0.381 & 0.349 & 0.332 & 0.321 & 0.315 & 0.310 & $\searrow$\\
$\bar{\Lambda}_{c}^{-}/D^{0}$ & 0.289 & 0.261 & 0.248 & 0.240 & 0.234 & 0.230 & $\searrow$\\
$R_{e/c\bar{c}}$ & 0.095 & 0.091 & 0.088 & 0.087 & 0.086 & 0.085 & $\searrow$\\
$R_{D^{0}/c\bar{c}}$ & 0.432 & 0.473 & 0.497 & 0.514 & 0.525 & 0.534 & $\nearrow$ \\
$R_{\mu/c\bar{c}}$ & 0.093 & 0.089 & 0.087 & 0.086 & 0.085 & 0.084 & $\searrow$\\
\hline
\end{tabular}
\label{tab:ratios-VP}
\end{table}

We further explore the charm ratio variances with the strangeness
suppression factor $\lambda_{s}$. The results are listed in Table
\ref{tab:ratios-lamds}. The ratios $D_{s}^{+}/D^{0}$ and
$D_{s}^{-}/D^{0}$ increase with $\lambda_{s}$ apparently. As the
total number of charm pairs is conserved while $\lambda_{s}$ is
changing, the more strange charm hadrons there are, and the fewer
nonstrange charm hadrons, so $R_{D^{0}/c\bar{c}}$ decreases on the
contrary. The increase of $ \lambda_{s} $ also results in the
relative reduction of the numbers of $u,d$ ($\bar{u}, \bar{d}$)
quarks. In the coalescence picture, the effect on the
$\Lambda_{c}^{+}$ ($udc$) or $\bar{\Lambda}_{c}^{-}$
($\bar{u}\bar{d}\bar{c}$) baryon from the number decrease of $u,d$
($\bar{u}, \bar{d}$) quarks is much larger than that on the $D^{0}$
($\bar{u}c$) meson, so the ratios $\Lambda_{c}^{+}/D^{0}$ and
$\bar{\Lambda}_{c}^{-}/D^{0}$ decrease with $\lambda_{s}$
increasing. The ratios $\frac{\bar{D}^{0}}{D^{0}}\sim
\frac{u}{\bar{u}}$ and $\frac{D^{-}}{D^{0}}\sim \frac{d}{\bar{u}}$
increase slightly, because the net baryon number or the net $u(d)$
quark number is invariant while the newly born
$u\bar{u}$($d\bar{d}$) pairs decrease. It is unexpected that
$R_{e/c\bar{c}}$ ($R_{\mu/c\bar{c}}$) is independent of
$\lambda_{s}$. This can be understood from Eq. (\ref{eq:Rec}), i.e.,
$D^{+}/D^{0}$, $\bar{D}^{0}/D^{0}$, and $D^{-}/D^{0}$ are nearly
constant, and the effects from $D_{s}^{+}/D^{0}$ ($D_{s}^{-}/D^{0}$)
and from $\Lambda_{c}^{+}/D^{0}$ ($\bar{\Lambda}_{c}^{-}/D^{0}$)
almost cancel out. One also can see that the error of
$R_{D^{0}/c\bar{c}}$ and the concomitant error of
$\sigma_{c\bar{c}}^{NN}$ from $\lambda_{s}$ is $\sim2\%$ and
$\sim4\%$ assuming the uncertainty of $\lambda_{s}$ is about $30\%$
due to the $\phi/K^{*0}$ data at RHIC \cite{Adler:2002sw,
Adams:2004ep}. Note that the error of $R_{D^{0}/c\bar{c}}$ from
$\lambda_{s}$ has no effect on the PHENIX $\sigma_{c\bar{c}}^{NN}$
from $R_{e/c\bar{c}}$.

\begin{table}
\caption{Same as Table \ref{tab:ratios-VP}, but with the parameter
$\lambda_{s}$.}
\begin{tabular}{c|c|c|c|c|c|c|c} \hline
$\lambda_{s}$ & 0.38 & 0.43 & 0.48 & 0.53 & 0.58 & 0.63 & \\
\hline
$D^{+}/D^{0}$ & 0.318 & 0.318 & 0.317 & 0.318 & 0.317 & 0.317 & \\
$D_{s}^{+}/D^{0}$ & 0.245 & 0.278 & 0.310 & 0.342 & 0.373 & 0.406 & $\nearrow$ \\
$\Lambda_{c}^{+}/D^{0}$ & 0.312 & 0.306 & 0.301 & 0.296 & 0.291 & 0.286 & $\searrow$ \\
$\bar{D}^{0}/D^{0}$ & 1.070 & 1.071 & 1.073 & 1.075 & 1.075 & 1.076 & \\
$D^{-}/D^{0}$ & 0.341 & 0.342 & 0.342 & 0.343 & 0.343 & 0.344 & \\
$D_{s}^{-}/D^{0}$ & 0.246 & 0.278 & 0.310 & 0.342 & 0.374 & 0.406 & $\nearrow$ \\
$\bar{\Lambda}_{c}^{-}/D^{0}$ & 0.239 & 0.235 & 0.230 & 0.225 & 0.221 & 0.217 & $\searrow$ \\
$R_{e/c\bar{c}}$ & 0.085 & 0.085 & 0.085 & 0.085 & 0.085 & 0.085 & \\
$R_{D^{0}/c\bar{c}}$ & 0.549 & 0.541 & 0.534 & 0.527 & 0.519 & 0.512 & $\searrow$ \\
$R_{\mu/c\bar{c}}$ & 0.084 & 0.084 & 0.084 & 0.084 & 0.084 & 0.084 & \\
\hline
\end{tabular}
\label{tab:ratios-lamds}
\end{table}

The dependencies on the net baryon number or $N_{\bar{d}}/N_{d}$ are
also investigated in the CO/RE mechanism. Note that the net baryon
number or the $u(d)$ quark number decreases with $N_{\bar{d}}/N_{d}$
increasing, so $\frac{D^{-}}{D^{0}}\sim \frac{d}{\bar{u}}$ and
$\frac{\bar{D}^{0}}{D^{0}}\sim \frac{u}{\bar{u}}$ decrease with
$N_{\bar{d}}/N_{d}$, and $\Lambda_{c}^{+}/D^{0}$
($\bar{\Lambda}_{c}^{-}/D^{0}$) decreases (increases) as the net
baryon number decreases. The ratios $\frac{D^{+}}{D^{0}} \sim
\frac{\bar{d}}{\bar{u}}$, $\frac{D_{s}^{+}}{D^{0}}\sim
\frac{\bar{s}}{\bar{u}}$, and $\frac{D_{s}^{-}}{D^{0}} \sim
\frac{s}{\bar{u}}$ are independent of $N_{\bar{d}}/N_{d}$. Based on
Eqs. (\ref{eq:Rec}) and (\ref{eq:Rdc}), the three key ratios
$R_{e/c\bar{c}}$, $R_{\mu/c\bar{c}}$, and $R_{D^{0}/c\bar{c}}$ are
also independent of the net baryon number as the sum of one kind of
charm hadron and its antiparticle is conserved, although the yield
of the kind of (anti)hadron varies with $N_{\bar{d}}/N_{d}$.

Next we discuss the effect from the charm quark $p_{T}$ spectrum.
Because the yields of charm hadrons are integrals of the hadronic
\pt spectra, their yields and ratios in the whole rapidity range
should be fixed with the input $N_{c\bar{c}}$ unchanged. However,
considering that the rapidity distributions of the secondary hadrons
from decay may be affected by the quark $p_{T}$ spectrum, the
midrapidity yields or ratios can also be influenced. To study the
effect quantitatively, keeping the input
$\sigma_{c\bar{c}}^{NN}=1.4$ mb and other parameters unchanged, we
replace only the charm quark spectrum by $f_{c}(p_{T}) =
(1.0+3.185p_{T}^{2})^{-1.0}/0.88$, which results in much harder
spectra of charm hadrons and nonphotonic leptons. These ratios,
including $R_{e/c\bar{c}}$, $R_{\mu/c\bar{c}}$, and
$R_{D^{0}/c\bar{c}}$, are all only slightly varied, i.e., they are
not sensitive to the quark spectrum. The charm hadron ratios are
then determined mainly by the CO/RE mechanism in QCM and some
corresponding parameters such as $V_{c}/P_{c}$ other than the \pt
spectra of quarks. Especially, with this property, if we assume the
CO/RE hadronization mechanism is universal in relativistic $AA$
reactions, we can give the prediction of charm hadron ratios by QCM
without the \pt spectra being well determined at LHC energy.

One sees that in the subsection, the ratio $R_{D^{0}/c\bar{c}}$
varies with $V_{c}/P_{c}$ and $\lambda_{s}$, while $R_{e/c\bar{c}}$
and $R_{\mu/c\bar{c}}$ are almost independent of all parameters
studied except for a very weak dependence on the parameter
$V_{c}/P_{c}$.

\subsection{Centrality dependencies of the ratios}

\begin{figure}
\includegraphics[scale=0.4]{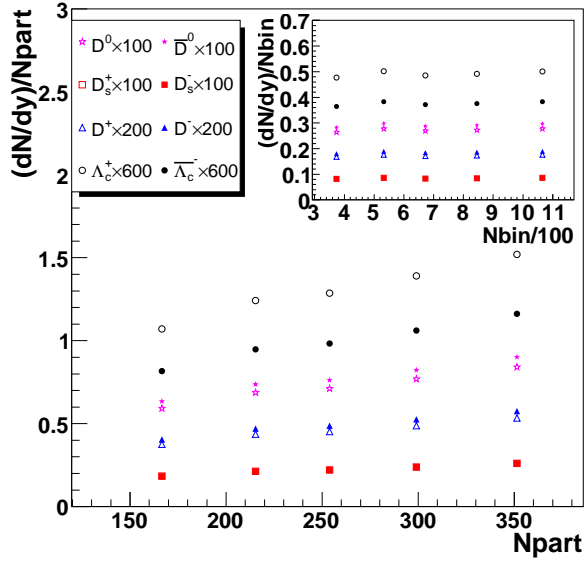}
\caption{(Color online) Centrality dependencies for midrapidity
densities of charm hadrons at 200 GeV.} \label{cap:density-Npart}
\end{figure}

We further investigate the centrality dependencies of these ratios
at 200 GeV in Au+Au collisions. According to the property of little
dependence on the charm quark $p_{T}$ spectrum, we can simply deal
with the charm quark spectra at various centrality bins by
neglecting the differences of radial flows. The numbers of charm
quarks at different centrality classes are given by Eq.
(\ref{eq:Ncc}). The parameter $N_{\bar{d}}/N_{d}$ does not vary with
the centrality based on the experimental observation of the ratios
of antiparticle to particle \cite{Adler:2003cb,:2008sb}. The other
parameters, such as $\lambda_{s}$, $D_{c}/O_{c}$, and $V_{c}/P_{c}$
are also assumed to be independent of centrality except for the
peripheral collisions. The centrality dependencies of rapidity
densities for charm hadrons are given in Fig.
\ref{cap:density-Npart}. One sees that the binary scaled rapidity
densities are all independent of $N_{bin}$. This is consistent with
Eq. (\ref{eq:Ncc}), as in CO/RE mechanism the multiplicity of the
single-charm hadron $M_{H_{c}}$ satisfies $M_{H_{c}} \propto
\frac{dM_{H_{c}}}{dy} \propto N_{c\bar{c}} \propto N_{bin}$. The
participant number scaled rapidity densities increase monotonically
with $N_{part}$ as we have $\frac{dM_{H_{c}}}{N_{part}dy} \propto
N_{part}^{1/3}$ based on Eq. (\ref{eq:Ncc}) and the approximate
relation $N_{bin} \propto N_{part}^{4/3}$. The result is consistent
qualitatively with that of the statistical hadronization model
\cite{Andronic:2003zv} although the input $N_{c\bar{c}}$ is
different. Note that the ratio of midrapidity density for
single-charm hadrons at a centrality
\begin{equation}\label{eq:R_Hc}
R_{H_{c}} = \frac{dM'_{H_{c}}/dy/N_{part}}{dM_{H_{c}}/dy/N_{part}} =
\frac{dM'_{H_{c}}/dy/N_{bin}}{dM_{H_{c}}/dy/N_{bin}} =
\frac{dM'_{H_{c}}/dy/N_{c\bar{c}}}{dM_{H_{c}}/dy/N_{c\bar{c}}}
\end{equation}
is irrelevant to the $N_{part}$, $N_{bin}$ or $N_{c\bar{c}}$, so the
ratios of single-charm hadrons and the three key ratios are all
independent of the centrality or $N_{c\bar{c}}$ as long as Eq.
(\ref{eq:Ncc}) is held and those parameters are independent of the
centrality.

\subsection{Energy dependencies of the ratios}

To explore the energy dependencies of the charm hadron ratios and
the three key ratios, furthermore, we study them at 130 and 62.4 GeV
center-of-mass energies at most central Au+Au collisions and 5.5 TeV
LHC energy for Pb+Pb collisions. $\sigma_{c\bar{c}}^{NN}=750$,  285
and about 20000 $\mu$b at the three energies are obtained from the
next-to-leading (NLO) pQCD calculations with $\mu_{R}$ equal to
$m_{c}$ \cite{Vogt:2001nh} (or see Fig. 1 in Ref. \cite{Xu:2006hk}).
We also use $\langle N_{bin} \rangle=965$ \cite{Adler:2002xw}, 904.3
\cite{Adams:2005cy} and 1303 (0-10\% centrality)
\cite{Armesto:2008fj}. Then we get $N_{c\bar{c}}\approx 18,$ 7, and
434 as inputs assuming $\sigma_{inel}^{pp}=41$, 36 and 60 mb,
respectively.

Note that Eq. (\ref{eq:R_Hc}) is still valid at different energies,
that is, within the CO/RE framework, the ratios of single-charm
hadrons and the three key ratios are independent of charm cross
section. This is important to the study of these ratios at various
energies, especially at LHC energy, as by now, the theory
predictions of $\sigma_{c\bar{c}}$ at LHC still have large
uncertainties \cite{Vogt:2001nh, Vogt:2004he, Zhang:2008zzc,
Liu:2008bw}. Assuming the CO/RE mechanism and the default values of
$D_{c}/O_{c}$ and $V_{c}/P_{c}$ are universal in relativistic heavy
ion collisions, the other factors that affect these ratios at
various energies are mainly $\lambda_{s}$ and the net baryon number.

Using $\lambda_{s} = 0.48$ and $\pbar/p = 0.98$
\cite{Eskola:2005ue}, the charm hadron ratios in top central Pb+Pb
collisions at 5.5 TeV are obtained. \footnotemark[5]
\footnotetext[5]{The results with input of $\pbar/p = 0.948$
\cite{Andronic:2005yp} are very close to those from $\pbar/p =
0.98$.} The LHC predictions, together with those at 130 and 62.4
GeV, are all listed in Table \ref{tab:densityratios}, where one can
see that $D_{s}^{+}/D^{0}$, $D_{s}^{-}/D^{0}$, $D_{s}^{+}/D^{+}$,
and $R_{D^{0}/c\bar{c}}$ vary mainly with $\lambda_{s}$ at different
energies. As the net baryon number decreases with the increasing
incident energy, the ratios $\Lambda_{c}^{+}/D^{0}$,
$\Sigma_{c}^{+}/D^{0}$, $\Sigma_{c}^{0}/D^{0}$, $\Xi_{c}^{+}/D^{0}$,
$\Xi_{c}^{0}/D^{0}$, $\bar{D}^{0}/D^{0}$, and $D^{-}/D^{0}$
decrease, while $\bar{\Lambda}_{c}^{-}/D^{0}$ and
$\bar{\Lambda}_{c}^{-}/\Lambda_{c}^{+}$ increase. The other ratios
$D^{+}/D^{0}$, $\Sigma_{c}^{+}/\Lambda_{c}^{+}$, $R_{e/c\bar{c}}$,
and $R_{\mu/c\bar{c}}$ are almost independent of the energy.

In this subsection, one can see that $R_{D^{0}/c\bar{c}}$ depends on
the incident energy due to its $\lambda_{s}$ dependency, whereas
$R_{e/c\bar{c}}$ and $R_{\mu/c\bar{c}}$ are independent of the
energy. The ratios $R_{e/c\bar{c}}$ and $R_{\mu/c\bar{c}}$ are
suitable quantities in relativistic heavy ion collisions for
determining the charm cross sections at different energies.

\section{Summary}

By using the QCM, we study the charm hadron production in top
central Au+Au collisions at $\sqrt{s_{NN}}=200$ GeV.  The $p_{T}$
dependencies of the charm baryon-to-meson ratios, such as
$\frac{\Lambda_{c}+\bar{\Lambda}_{c}}{D^{0}+\bar{D}^{0}}$,
$\frac{\Lambda_{c}+\bar{\Lambda}_{c}}{D^{+}+D^{-}}$, and
$\frac{\Lambda_{c}+\bar{\Lambda}_{c}}{D_{s}^{+}+D_{s}^{-}}$ are
obtained. One can see that the charm baryon enhancement in
intermediate $p_{T}$ range, similar to that of the p/$\pi$ ratio, is
very prominent. The shape differences from that of the p/$\pi$ ratio
are mainly from the hard charm quark spectrum. The $R_{AA}$ ordering
with strangeness content for charm hadrons is discussed, and it can
be regarded as a powerful signal for proving the CO/RE hadronization
mechanism for open charm hadrons. The midrapidity densities of the
single-charm hadrons and their ratios at 200 GeV are calculated.
These ratios in $AA$ collisions, including the key ratios
$R_{e/c\bar{c}}$ and $R_{\mu/c\bar{c}}$, are apparently different
from those in $pp(\bar{p})$ reactions or $e^{+}e^{-}$ annihilations
because of the charm baryon enhancement and the strangeness
enhancement. With the newest branching ratios from PDG, the modified
charm hadron ratios lead to a $\sim17\%$ increase of the central
value of $\sigma_{c\bar{c}}^{NN}$ measured by PHENIX. However, this
correction is not enough to account for the discrepancy of
$\sigma_{c\bar{c}}^{NN}$ between STAR and PHENIX. Considering the
uncertainties of some parameters, we systematically explore the
charm ratio dependencies on various parameters. Assuming the CO/RE
hadronization mechanism is universal in relativistic heavy ion
collisions, we further investigate the energy dependencies of these
ratios, and reveal that these ratios are mainly dependent on
$\lambda_{s}$ and the net baryon number but not on the charm cross
section. The predictions of charm hadron ratios at $\sqrt{s_{NN}} =
62.4, 130$, and 200 GeV for the upgrade of RHIC and at 5.5 TeV for
LHC are given. These ratios are important for the precise
measurement of $\sigma_{c\bar{c}}^{NN}$ in the future, and the CO/RE
hadronization mechanism for charm hadrons can be tested at RHIC and
LHC.

\acknowledgments

The authors thank Lie-Wen Chen, Che-Ming Ko, Fu-Qiang Wang, Qun
Wang, and the colleagues of THPP in Shandong University for helpful
discussions. Special thanks go to Chen Zhong (SIAP) for rendering
some data adopted in the manuscript. The work is supported in part
by National Natural Science Foundation of China (NSFC) under grant
Nos. 10475049, 10775089, 10775090.


\end{document}